\documentclass[12pt,a4paper,final]{iopart}

\usepackage{footnote}
\usepackage{textcomp}
\usepackage{color}
\usepackage{multicol}
\usepackage{bm}         
\usepackage{ulem}
\usepackage{iopams}  
\usepackage{graphicx}
\usepackage[breaklinks=true,colorlinks=true,linkcolor=blue,urlcolor=blue,citecolor=blue]{hyperref}

\begin{document}

\title{Covariance analysis for Energy Density Functionals and instabilities}

\author{X. Roca-Maza$^{1}$, N. Paar$^{2}$ and G. Col\`o$^{1}$}
\address{
$^1$Dipartimento di Fisica, Universit\`a degli Studi di Milano and INFN, Sezione di Milano, 20133 Milano, Italy\\
$^2$Department of Physics, Faculty of Science, University of Zagreb, Zagreb, Croatia
}
\eads{\mailto{xavier.roca.maza@mi.infn.it}}

\begin{abstract}
We present the covariance analysis of two successful nuclear energy density functionals, (i) a non-relativistic Skyrme functional built from a zero-range effective interaction, and (ii) a relativistic nuclear energy density functional based on density dependent meson-nucleon couplings. The covariance analysis is a useful tool for understanding the limitations of a model, the correlations between observables and the statistical errors. We show, for our selected test nucleus ${}^{208}$Pb, that when the constraint on a property $A$ included in the fit is relaxed, correlations with other observables $B$ become larger; on the other hand, when a strong constraint is imposed on $A$, the correlations with other properties become very small. We also provide a brief review, partly connected with the covariance analysis, of some instabilities displayed by several energy density functionals currently used in nuclear physics.  
\end{abstract}
%
%
%
\vspace{2pc}

\submitto{\JPG}

\section{Introduction}

A successful methodology for an effective description of nuclei along the periodic table corresponds to the self-consistent mean field approach. This class of models can be understood as an approximate realisation of a nuclear energy density functional (EDF). The Density Functional Theory is a powerful and general approach used successfully in physics, chemistry and material science \cite{Martin}. In condensed matter, it is possible to describe with an exquisite accuracy many-electron systems, though with some exceptions. To some extent, the reason relies on the possibility of deriving such functionals from {\it ab-initio} calculations of the electron gas. At present, calculations based on the use of realistic nucleon-nucleon interactions in the vacuum lacks sufficient accuracy for the description of ground and excited state properties of medium to heavy mass nuclei \cite{hofmann01, hirose07,soma13,ab1,ab2,ab3,ab4,ab5,ab6,ab7,ogawa11}. Instead, one can derive accurate effective interactions characterised by a relatively small number of parameters to be adjusted --- on the order of, or less than, ten.

In general, most of the nuclear effective models available in the literature omit theoretical error estimations. This leads to optimal model parametrisation with respect to a given quality measure --- as for example a $\chi^2$ --- of limited use for some extrapolations. Not assessing the errors in the determination of the parameters of a given model may lead to unreliable conclusions when extrapolating far away from nuclei used in the $\chi^2$ minimisation.

When proposing a nuclear structure model, one usually formulate it in terms of a minimal number of effective interaction terms and associated parameters. Although adding more parameters to any existing reasonable model may improve the quality measure it does not necessarily mean that the overall quality of the fit will be improved. For instance, a non desirable consequence of adding a new parameter can be that large changes are produced in the already existing parameters. This is a clear signature that the model with the new parameter is introducing uncontrolled correlations, and that it may have converged to a local (i.e. not to a global) minimum. When this occurs, the confidence intervals and standard deviations predicted for all (or some of) the parameters suffer a large increase. Also, flat minima are sign of redundancies on the parameters. That is, rather different values of one (or more) parameters produce slight changes in the quality measure, and this is a common fingerprint of an over-parametrised model.

There are several strategies one can follow in order to deal with this problem. The approach based on the covariance analysis allows to determine first if the model contains redundant parameters and possibly, then, to identify which one can or should be removed, or {\it a priori} fixed, due to physical considerations. Although intimately related, we will not focus on the latter issue. We would like to focus on describing a strategy through which existing accurate models can help us in finding key observables and nuclei.  This is possible via the covariance analysis of a given model \cite{Kortelainen:2010, reinhard10, Fattoyev:2011, Fattoyev:2012, Piekarewicz:2012, Reinhard:2013a, Erler:2013, Reinhard:2013b, dobaczewski14} because it provides a measure of the statistical uncertainties of an observable, as well as the correlations between observables, on the basis of the experimental data used for defining the quality measure. We will also devote part of this manuscript to briefly describe the appearance of some instabilities in the functionals. Although a connection has not been proven yet, instabilities should be a warning, when building a new functional, that complements the aforementioned signals of unreliability based on covariance analysis. 

There exists another type of uncertainty associated to the choice of the model: the {\it systematic uncertainty}. In the absence of an exact model for finite nuclei, a possible way to estimate the model dependence, i.e. systematic uncertainty, associated to a given prediction is to compare different kinds of EDFs. In Ref.~\cite{erler12}, both statistical and systematic uncertainties have been investigated within a selected set of non-relativistic EDFs. It has been found that systematic uncertainties govern the uncertainty in extrapolated mass differences. This feature might also be true for their relativistic counterparts \cite{agbemava14}, although this has not been studied in detail yet. We will not discuss systematic uncertainties in the present work since the scope of this contribution is to focus on the statistic uncertainties. 

The manuscript is organised as follows. In Sec.~\ref{theory}, we present a brief review of the technique to perform the covariance analysis \cite{bevington,dobaczewski14}. In Sec.~\ref{edfs}, we present the used functionals and give some details on the definition of the corresponding quality functions employed in the fitting protocol. In Sec.~\ref{results}, we will provide the results of the 
covariance analysis for each of the models. In Sec.~\ref{instabilities} we discuss some of the instabilities associated with the energy density functionals of current use in nuclear physics. Our conclusions and outlook are laid in Sec.~\ref{conclusions}.  

\section{Covariance analysis}
\label{theory}

A brief review on the method of covariance analysis and some useful technical details are given in this Section. Many textbooks contain a more exhaustive treatment of the formalism. Here, we just refer the interested reader to a general book \cite{bevington}, and to the most recent work devoted to its application in the nuclear case \cite{dobaczewski14}. 

\subsection{Formalism}  
\label{formalism}
Consider a model characterised by $n$ parameters $\bm{p} = (p_1 , . . . , p_n)$. Those parameters define the model space and can be coupling constants of an effective interaction. Observables ($\mathcal{O}$) are, therefore, functions of the parameters $\mathcal{O}(\bm{p})$.

\subsubsection{$\chi^2$ definition}
The $\chi^2$ defines here the quality measure. It reads
\begin{equation}
\chi^2(\bm{p}) = \sum_{\imath=1}^{m}
 \left(\frac{\mathcal{O}_\imath^{\rm theo.}(\bm{p})-\mathcal{O}_\imath^{\rm ref.}}
 {\Delta\mathcal{O}_\imath^{\rm ref.}}\right)^2 \,
\label{chi2}
\end{equation}
where ``theo.'' stands for the calculated values, and ``ref.'' may refer to experimental, observational and/or {\it pseudo-data}\footnote[1]{{\it pseudo-data} correspond to a derived quantity, not directly observable, that is sometimes used in the definition of the quality measure as a benchmark.} that sometimes are used to guide the models. The use of {\it pseudo-data} should be taken with care, as we will discuss with an example in Sec.~\ref{results}. $\Delta\mathcal{O}_\imath^{\rm ref.}$ are the {\it adopted errors} that, strictly speaking, should stand for the experimental standard deviations. This choice is not always reasonable as in some cases the experimental error may be smaller than the intrinsic accuracy of the fitted functional, and a small $\Delta\mathcal{O}_\imath^{\rm ref.}$ may prevent the fitting protocol from converging. In summary, some freedom exists in choosing a convenient set of $\mathcal{O}_\imath^{\rm ref.}$ and $\Delta\mathcal{O}_\imath^{\rm ref.}$ that with little redundancies may characterise the nucleus. 

\subsubsection{Covariance analysis of parameters and observables}

Assuming that the $\chi^2$ is a well behaved, analytical hyper-function of the parameters around their optimal value $\bm{p}_0$, $\partial_{\bm{p}}\chi^2(\bm{p})\mid_{\bm{p}=\bm{p}_0} = 0$, and that the $\chi^2$ near the minimum can be approximated by a Taylor expansion as a hyper-parabola in the parameter space, we can write
\begin{equation}
\chi^2(\bm{p})-\chi^2(\bm{p}_0) \approx 
 \frac{1}{2}\sum_{\imath, \jmath}^n(p_{\imath}-p_{0\imath})
 \partial_{p_\imath}\partial_{p_\jmath}\chi^2\vert_{\bm{p}_0}(p_{\jmath}-p_{0\jmath})\ \ .
\label{curvature}
\end{equation}
This expression defines the curvature matrix, $\mathcal{M}\equiv\partial_{p_\imath}\partial_{p_\jmath}\chi^2\vert_{\bm{p}_0}$. $\mathcal{M}$ provides access to estimate the errors ($\bm{e}$) of the fitted parameters as follows, 
\begin{equation}
e_\imath \equiv e(p_\imath) = 
 \sqrt{\left(\mathcal{M}^{-1}\right)_{\imath\imath}} \equiv 
 \sqrt{\mathcal{E}_{\imath\imath}} \ \ , 
\label{e}
\end{equation}
where we have defined the covariance (or error) matrix $\mathcal{E}$. The meaning of this definition for the error in the parameters can be qualitatively understood as follows. If the curvature matrix takes a large (small) value along the $p_i$ direction, it means that a given change in the parameter $\Delta p_\imath$ will produce a large (small) change in $\chi^2(\bm{p}_0 + \Delta p_\imath) - \chi^2(\bm{p}_0)$. Therefore, the parameter $p_\imath$ will (will not) be accurately determined and its error $\sqrt{\mathcal{M}_{\imath\imath}^{-1}}$ will be small (large). The equation $\chi^2(\bm{p}_0 + \Delta p_\imath) - \chi^2(\bm{p}_0)=1$ is used to define the magnitude of $\Delta p_\imath$. 

The covariance or error matrix can be further exploited and also the correlation matrix ($\mathcal{C}$) can be estimated,  
\begin{equation}
\mathcal{C}_{\imath\jmath} \equiv \frac{\mathcal{E}_{\imath\jmath}}
 {\sqrt{\mathcal{E}_{\imath\imath}\mathcal{E}_{\jmath\jmath}}} 
\label{c}
\end{equation}
where $\mathcal{C}_{\imath\jmath}$ takes values form $-1$ to $1$. $\mathcal{C}_{\imath\jmath}\approx 1$ indicates a large correlation and $-1$ a large anti-correlation between parameters $p_\imath$ and $p_\jmath$, respectively. This would indicate that $p_\imath$ (or $p_\jmath$) is redundant and can be fixed during the fit by setting its value to a physically reasonable value -- or to zero if needed. On the contrary, $\mathcal{C}_{\imath\jmath}$ around zero means that no correlation holds at all between parameters $p_\imath$ and $p_\jmath$. This clearly indicates that both parameters are needed for the description of the set of observables used for the fit.

Moreover, once the set of parameters minimising the $\chi^2$ have been determined, the expectation value of an observable $A$, not included in the fit, can be computed as $A(\bm{p}_0)$. The uncertainties in the prediction of such observable are originated by the {\it adopted errors} in the fitted observables. To estimate such an error --- its {\it adopted-standard deviation} in a sense --- one can expand the observable under study, $A(\bm{p})$, around the minimum $\bm{p}_0$ assuming a smooth behaviour and neglecting second and higher order derivatives,     
\begin{equation}
A(\bm{p}) = A(\bm{p}_0) + (\bm{p}-\bm{p}_0)
\partial_{\bm{p}}A(\bm{p})\mid_{\bm{p}=\bm{p}_0} \ .
\label{obs}
\end{equation}
Within this approximation the statistical expectation value of the observable $\overline{A}$ would coincide exactly with $A_0$\footnote[2]{This can be demonstrated if we assume a Gaussian distribution of the different parametrisations around the minimum, $\mathcal{P}(\bm{p}) = \mathcal{N}\exp\left(-\frac{1}{2} (\bm{p}-\bm{p}_0)\mathcal{M}(\bm{p}-\bm{p}_0)\right)$, where $\mathcal{N}$ is a normalisation constant.}. From here, one can calculate the covariance between two observables by using Eq.~(\ref{obs}) and within the adopted approximations as, 
\begin{equation}
C_{AB}       = \overline{(A(\bm{p})-\overline{A}) (B(\bm{p})-\overline{B})} 
\approx \sum_{\imath\jmath}^n 
\left.\frac{\partial A(\bm{p})}{\partial p_\imath}\right\vert_{\bm{p}=\bm{p}_0}
\mathcal{E}_{\imath\jmath}
\left.\frac{\partial B(\bm{p})}{\partial p_\jmath}\right\vert_{\bm{p}=\bm{p}_0}\ \ .
\label{co}
\end{equation}
The variance of $A$ which estimates the uncertainty of this observable is, then, easily calculated from the latter expression as $C_{AA}$. Furthermore, one may also calculate the Pearson-product moment correlation coefficient between those observables,           
\begin{equation}
c_{AB} \equiv \frac{C_{AB}}{\sqrt{C_{AA}C_{BB}}} \ \ , 
\label{corr-pearson}
\end{equation}
a quantity very useful in the analysis of correlations between predicted observables, that will be used along the present work. In analogy with the correlation coefficient defined in Eq.~(\ref{c}), $c_{AB}=1$ means complete correlation between observables $A$ and $B$, whereas $-1$ means complete anti-correlation and $c_{AB}=0$ means no correlation at all.

\subsection{Numerical details}
\label{details}
Here we will briefly give some details we think might be useful for the reader. 

\subsubsection{The curvature matrix $\mathcal{M}$}   
The calculation of the curvature matrix, proportional to the Hessian matrix, can be done by using different numerical approximations. In the present study, we have followed the assumptions used along Sec.~\ref{formalism}. In this case, one can calculate the curvature matrix starting form Eq.~(\ref{chi2}) in a simplified and numerically convenient way (see \cite{bevington}) as follows, 
\begin{equation}
\partial_{p_\imath}\partial_{p_\jmath}\chi^2(\bm{p}) \approx2\sum_{k=1}^{m}\frac{\partial_{p_\imath} \mathcal{O}_k^{\rm theo.}(\bm{p})}
 {\Delta\mathcal{O}_k^{\rm ref.}}\frac{\partial_{p_\jmath} \mathcal{O}_k^{\rm theo.}(\bm{p})}
 {\Delta\mathcal{O}_k^{\rm ref.}} 
\label{curvature-2}
\end{equation}
and then, only first derivatives should be calculated\footnote[3]{We have employed a symmetric two point formula for performing the first derivatives since they are accurate in describing smooth behaviours and their associated error is proportional to $\partial_{p_\imath}^3\mathcal{O}$.}.

\subsubsection{How to chose step sizes for calculating derivatives with respect to the parameters}
The region of reasonable parametrisations is enclosed by the contour $\chi^2(\bm{p})-\chi^2(\bm{p}_0)\approx 1$, since this ensures that (on average) the steps in the parameters provide a change comparable to the {\it adopted errors}. For this reason, a reasonable choice for the step size is such that the variation in each parameter produces a change $\Delta\chi^2\approx 1$. Assuming a parabolic approximation of $\chi^2$ around the minimum and using Eq.~(\ref{e}) the estimate of $\Delta p_\imath$ is the following,    
\begin{equation}
e(p_\imath)^2 \equiv\left(\Delta p_\imath\right)^2 = 
 \left(\mathcal{M}^{-1}\right)_{\imath\imath} \equiv 
 2\left(\left.\frac{\partial^2\chi^2}{\partial p_\imath^2}\right\vert_{\bm{p}_0}\right)^{-1} \ .
\label{step}
\end{equation}
We have also checked that increasing or decreasing artificially the value of $\Delta p_\imath$ as calculated by using Eq.~(\ref{step}), lead us to similar results for the curvature and covariance matrices.   

\section{Energy density functionals}
\label{edfs}
In this Section, we present the non-relativistic and relativistic functionals used in the present analysis, including relevant references that completely define their functional form and the corresponding merit functions employed in the fitting protocol.

\subsection{Skyrme energy density functional}
\label{skyrme}
The $\chi^2$ associated to the Skyrme functional named SLy5-min has been defined as similarly as possible to the one used for the successful SLy5 functional \cite{chabanat1997,chabanat1998}. We will also present two variants of SLy5-min produced by slightly changing the $\chi^2$ definition. This exercise will be very useful in showing the impact of the {\it adopted errors}, and of using {\it pseudo-data}, on the correlations between different observables. For all the details in the definition of the $\chi^2$, fitting procedure, values of the parameters and properties of SLy5, see Refs.~\cite{chabanat1997,chabanat1998}. In analogy with the original fitting protocol of SLy5, we have fixed the spin-orbit parameters ($W_0 = W_0^\prime = 126$ MeV fm${}^{5}$), one of the parameters in the attractive part of the interaction ($x_2=-1$) as well as the parameter controlling the density dependent part of the effective interaction ($\alpha=1/6$). 

The $\chi^2$ used for fitting SLy5 and SLy5-min is defined in Eq.~(4.1) of Ref.~\cite{chabanat1997}. Specifically, it includes the binding energies of ${}^{40,48}$Ca, ${}^{56}$Ni, ${}^{130,132}$Sn and ${}^{208}$Pb with a fixed adopted error of 2 MeV, the charge radius of ${}^{40,48}$Ca, ${}^{56}$Ni and ${}^{208}$Pb with a fixed adopted error of 0.02 fm, the neutron matter Equation of State calculated by Wiringa {\it et al.} in Ref.~\cite{wiringa} for densities between 0.07 and 0.40 fm${}^{-3}$ with an adopted error of 10\%, and the saturation energy ($e(\rho_0) = -16.0 \pm 0.2$ MeV) and density ($\rho_0=0.160\pm 0.005$ fm${}^{-3}$) of symmetric nuclear matter. We adopted the same values for the experimental data as taken in \cite{chabanat1997}. Although nuclear matter properties are not real laboratory data, it is fair to state that at the moment when SLy5 has been proposed, the calculations by Wiringa and collaborators were considered as state-of-the-art; the Lyon group wanted to be able to extrapolate SLy* forces to describe neutron star matter ($\rho \sim 2-3\rho_0$) where we do not have precise information at our disposal. Also, the adopted values and errors for the saturation energy and density of symmetric nuclear matter are still nowadays widely accepted. So, {\it pseudo-data} may help in guiding nuclear models and foster new advances in the field \cite{tamara08, baldo13, xrm11}. The values of optimal parameters $\bm{p}_0$ and associated statistical errors $\sqrt{\mathcal{E}_{ii}}$ for SLy5-min\footnote[4]{The differences in the parameters with respect to the original SLy5 range from few \textperthousand ~to a few \%~ except for the $x_1$ parameter. These discrepancies are mainly due to the slightly different fitting protocols used in the optimization of SLy5 and SLy5-min. The main difference is that we do not fix the value of the isovector dipole enhancement factor $\kappa$ in SLy5-min while in SLy5 it was fixed.} functional are shown in Tab.~\ref{tabl1}.

\Table{
\label{tabl1}
Parameter name $\bm{p}$, their optimal value $\bm{p}_0$, and deviation $\sqrt{\mathcal{E}_{ii}}$ for SLy5-min and for DDME-min1. 
}
\br
\multicolumn{4}{l}{SLy5-min}  & &\multicolumn{5}{l}{DDME-min1}  \\
\cline{1-4} \cline{6-9}
  $\bm{p}$ & $\bm{p}_0$ & &$\sqrt{\mathcal{E}_{ii}}$& units&  $\bm{p}$ & $\bm{p}_0$ & &$\sqrt{\mathcal{E}_{ii}}$& units\\          
\mr                                                                                                                        
  $t_0$      &  $-2475.408$ &$\pm$ &$149.455$& MeV fm$^3$& $m_{\sigma}$      &  $549.841$ &$\pm$ &$ 1.988$& MeV\\            
  $t_1$      &  $482.842$ &$\pm$   &$58.537$& MeV fm$^5$&  $m_{\omega}$      &  $783.000$ &  &fixed & MeV\\    
  $t_2$      &  $-559.374$ &$\pm$  &$144.534$& MeV fm$^5$&  $m_{\rho}$      &  $763.000$ & &fixed & MeV\\                   
  $t_3$      &  $13697.07$ &$\pm$  &$1672.93$& MeV fm$^{3+3\alpha}$&  $g_{\sigma}(\rho_{sat})$      &  $10.544$ &$\pm$ &$0.144  $& \\ 
  $x_0$      &  $0.741185$ &$\pm$  &$0.189191$& &    $g_{\omega}(\rho_{sat})$      &  $13.031$ &$\pm$ &$0.170$& \\    
  $x_1$      &  $-0.146374$ &$\pm$ &$0.468173$& &   $g_{\rho}(\rho_{sat})$      &  $3.798$ &$\pm$ &$0.247$& \\       
  $x_2$      &  $-1$ &      &fixed &          &            $b_{\sigma}$      &  $1.117$ &$\pm$ &$0.590$& \\                   
  $x_3$      &  $1.162688$ &$\pm$  &$0.340537$& &     $c_{\sigma}$      &  $1.676$ &$\pm$ &$0.948$& \\                 
  $W_0$      &  $126$   &      &fixed & MeV fm$^5$&       $b_{\omega}$      &  $0.934$& $\pm$ &$0.628$ &   \\                 
  $W_0^\prime$&   $126$  &      &fixed & MeV fm$^5$&       $c_{\omega}$      &  $1.411$ &$\pm$ &$1.034$& \\                      
  $\alpha$   &  $1/6$  &      &fixed &  &      $a_{\rho}$    &  $0.524$ & $\pm$ & $0.194$ &  \\                   
\br
\end{tabular}
\end{indented}
\end{table}

\subsection{Covariant energy density functional}
The formulation of the relativistic nuclear energy density functional with density dependent meson-nucleon couplings is based on Ref.~\cite{Nik.02}. For the purpose of the present study we employ the DDME-min1 functional based on an effective finite-range interaction. More details about the theoretical framework and its implementation are given in Refs.~\cite{PVKC.07,Nik.14}.

The optimal parametrisation DDME-min1 is obtained by a $\chi^2$ minimisation using ground state properties of 17 even-even spherical nuclei, ${}^{16}$O, ${}^{40,48}$Ca, ${}^{56,58}$Ni, ${}^{88}$Sr, ${}^{90}$Zr, ${}^{100,112,120,124,132}$Sn, ${}^{136}$Xe, ${}^{144}$Sm and ${}^{202,208,214}$Pb (same set as in Ref.~\cite{Bue.02}). Specifically, the properties included are nuclear binding energies, charge radii, diffraction radii and surface thicknesses (for definitions see Ref.~\cite{Bue.02}). In the least squares fit, the assumed errors of these observables are 0.2\%, 0.5\%, 0.5\%, and 1.5\%, respectively. For open shell nuclei a BCS approach is adopted. The neutron and proton pairing gaps are fixed to be equal to the mass differences of neighbouring nuclei by using a five point formula. The values of optimal parameters $\bm{p}_0$ and respective deviations $\sqrt{\mathcal{E}_{ii}}$ for DDME-min1 functional are shown in Tab.~\ref{tabl1}.

\section{Results}
\label{results}
In this section we discuss the main results obtained via the covariance analysis of the two successful EDFs described in Sec.~\ref{edfs}. As an example, we will also present a study of the sensitivity of our results when employing the SLy5-min functional when (i) the weight in the neutron matter equation of state is relaxed and (ii) when the respective weight is further relaxed and the neutron skin thickness in ${}^{208}$Pb is added into the $\chi^2$ definition with a very large weight, i.e. small adopted error.   

\subsection{SLy5-min and DDME-min1} 
\label{results1}

\begin{figure}[t!]
\begin{center}
\includegraphics[clip=true,width=0.48\linewidth]{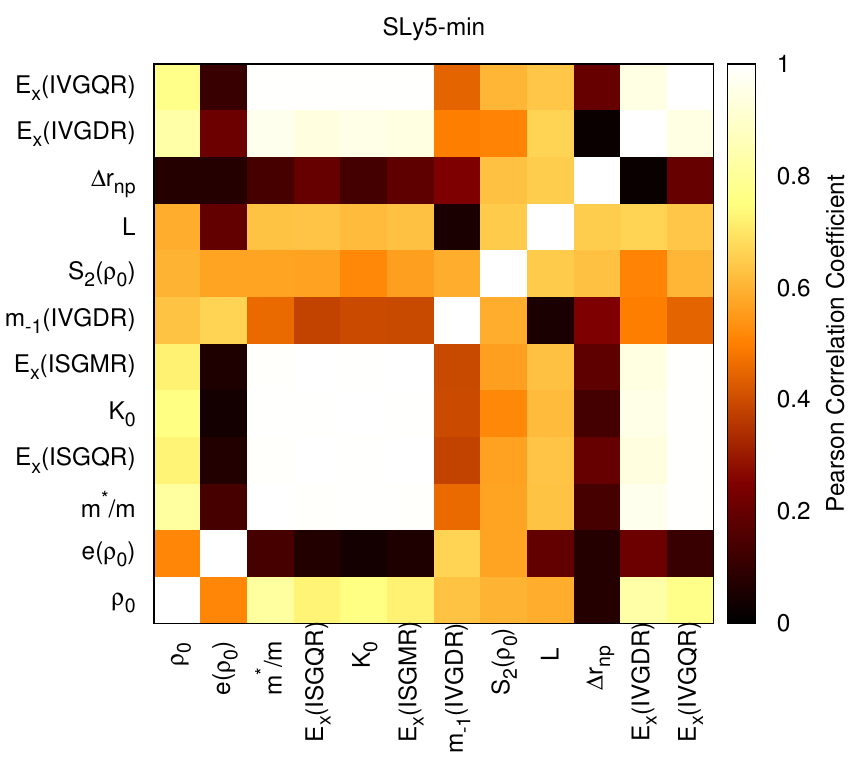}\hfill
\includegraphics[clip=true,width=0.48\linewidth]{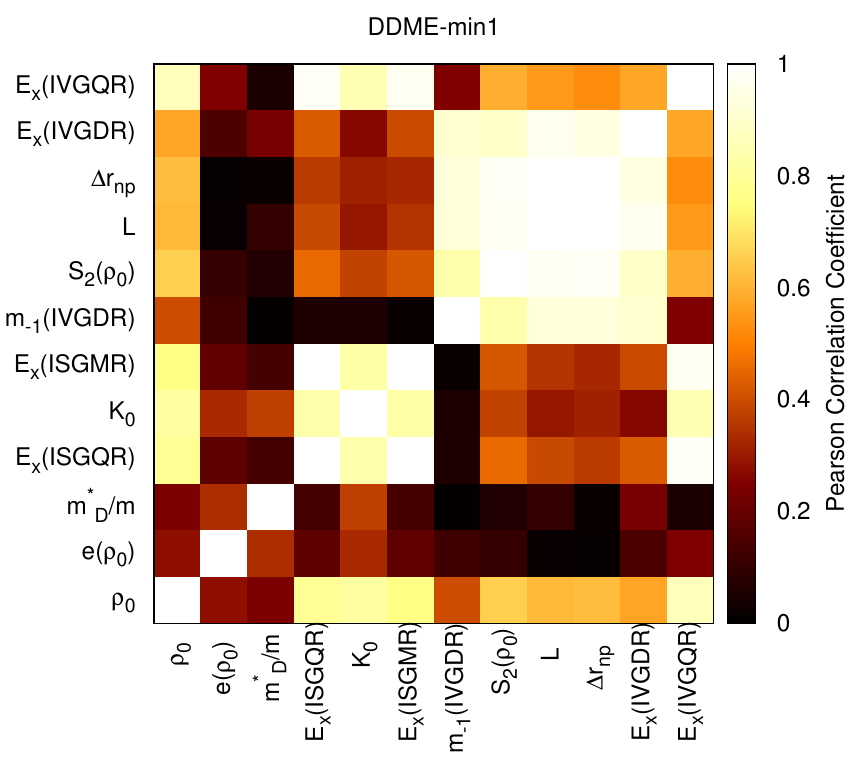}
\caption{Pearson product-moment correlation coefficient matrix (colour code) as predicted by the covariance analysis based on SLy5-min (left panel) and DDME-min1 (right panel) for various properties of nuclear matter and $^{208}$Pb (see text for the definition).\label{fig01}}
\end{center}
\end{figure}

In Fig.~\ref{fig01}, we have depicted the absolute value of the Pearson product-moment correlation coefficient matrix (colour code) as predicted by the covariance analysis of SLy5-min and DDME-min1 for some well known properties\footnote[5]{All the calculated properties in nuclei refer to ${}^{208}$Pb.} that serve us as an example (vertical axis from top to bottom): $E_x$(IVGQR) centroid energy of the Isovector Giant Quadrupole Resonance; $E_x$(IVGDR) centroid energy of the Isovector Giant Dipole Resonance; $\Delta r_{np} \equiv r_n - r_p$ neutron skin thickness; $L$ slope parameter of the symmetry energy at saturation density $L=3\rho_0 \partial_\rho S_2(\rho)\vert_{\rho_0}$; $S_2(\rho_0)$ symmetry energy at saturation; $m_{-1}$(IVGDR) inverse energy weighted sum rule of the Isovector Giant Dipole Resonance; $E_x$(ISGMR) centroid excitation energy of the Isoscalar Giant Monopole Resonance; $K_0$ nuclear matter incompressibility $K_0=9\rho_0^2 \partial_\rho^2 e(\rho)\vert_{\rho_0}$; $E_x$(ISGQR) centroid energy of the Isoscalar Giant Quadrupole Resonance; $m^*/m$ ($m^*_D/m$) nuclear matter Schr\"odinger (Dirac) effective mass divided by the nucleon mass $m$; $e(\rho_0)$ nuclear matter saturation energy; $\rho_0$ nuclear matter saturation density. Note that the matrix is symmetric.

The main features to be discussed in Fig.~\ref{fig01} are the following. First of all, the strong correlation between the isoscalar properties: $E_x$(ISGMR) in ${}^{208}$Pb, $K_0$, $E_x$(ISGQR) in ${}^{208}$Pb and $m^*/m$\footnote[6]{Note that for the DDME-min1 functional, the Dirac effective mass is not strongly correlated with any other property opposite to what happens with the SLy5-min predictions for the Schr\"odinger effective mass. This is probably due to the different nature of the Dirac effective mass \cite{jaminon89}.}. This might be expected in general due to their common isoscalar nature. Note that $\rho_0$ and $e(\rho_0)$ are also isoscalar properties, but while the former is still correlated with the previously mentioned properties as well as with $E_x$(IVGQR) \cite{XRM.13}, the latter seems to be uncorrelated with them. Only for the SLy5-min results, the saturation energy of nuclear matter $e(\rho_0)$ is correlated with $\rho_0$, $m_{-1}$(IVGDR) and $S_2(\rho_0)$. Such a correlation can be understood from the model relation that holds between them in neutron matter ($\delta \equiv (\rho_n - \rho_p) / \rho = 1$) explicitly constrained in the fit\footnote[7]{The $\chi^2$ of DDME-min1 does not contain such information.}. That is, $e(\rho_0,\delta=1) \approx e(\rho_0) + S_2(\rho_0)$. So, here it is clear that also correlations between isoscalar and isovector properties may arise depending on the definition of the $\chi^2$ --- cf. the left panel of Fig.~\ref{fig01}. On the contrary, for the case of DDME-min1, isoscalar quantities weakly correlate with the observables of isovector character. Nevertheless, as an exception, $E_x$(IVGQR) is highly correlated with the properties of isoscalar giant resonances within both models. This can be understood in terms of both macroscopic and microscopic models \cite{XRM.13}.

Isovector properties show a clear mutual correlation in both models, though it is higher for the relativistic functional. However, when considered in more detail, SLy5-min shows some (apparently) puzzling features. $\Delta r_{np}$ is not predicted to be correlated neither with $E_x$(IVGQR) nor with $E_x$(IVGDR). Actually, it has been shown in both cases that $\Delta r_{np}$ has a non-linear dependence on other quantities as well. Such a dependence may prevent an approximate linear correlation. In the former case, $\Delta r_{np}$ is basically related with a combination of $E_x$(IVGQR), $E_x$(ISGQR) and $S_2(\rho_0)$ \cite{XRM.13}, and in the latter case with $E_x$(IVGDR), $S_2(\rho_0)$ and the isovector dipole enhancement factor $\kappa$ \cite{trippa08}. 

On top of that, $m_{-1}$(IVGDR) is neither predicted to be correlated with $\Delta r_{np}$ nor with $L$ by the SLy5-min functional. Such a behaviour is in agreement with the analysis of a large set of EDFs guided by a Droplet Model based formula for $m_{-1}$(IVGDR) \cite{roca-maza13b}. Specifically, this formula shows that $m_{-1}$(IVGDR) depends on $L$ (or $\Delta r_{np}$) and also on other quantities such as $S_2(\rho_0)$ in a non-linear way (cf. Eq.~(8) of Ref.~\cite{roca-maza13b}).       
 
Some of these results are in agreement with previous covariance analysis performed for different functionals \cite{Kortelainen:2010, reinhard10, Fattoyev:2011, Fattoyev:2012, Piekarewicz:2012, Reinhard:2013a, Erler:2013, Reinhard:2013b, dobaczewski14}. Specifically, the correlations between the $\Delta r_{np}$ and isovector quantities such as $L$, $m_{-1}$(IVGDR) or $S_2(\rho_0)$ and the lack of correlation between the $\Delta r_{np}$ and isoscalar quantities such as $m^*/m$, $K_0$, $\rho_0$, $E_x$(ISGMR) or $E_x$(ISGQR) are common  in some of the analysis --for clarity, we recall here that all the correlations in properties of finite nuclei are referred to ${}^{208}$Pb.    

The analysis of correlations predicted by DDME-min1 and SLy5-min provides an indication for possible constraints for $L$ and $S_2(\rho_0)$ using the information on isovector giant resonances. An important aspect of this analysis regarding the DDME-min1 functional are also the strong correlations obtained between $\Delta r_{np}$, $E_x$ of IVGDR and IVGQR, as well as dipole polarizability (proportional to $m_{-1}$(IVGDR)). The somewhat different outcome from SLy5-min will be further studied in the next Sec.~\ref{results2}.  
 
\Table{
\label{tabl2}
Mean values and deviations of the different properties, $A$, used for the calculation of the Pearson-product correlation coefficient as predicted by SLy5-min and DDME-min1. The first half of the table refers to infinite symmetric nuclear matter properties (SNM) and the second one to properties of ${}^{208}$Pb. Note that $m^*/m$ stands for the Dirac effective mass in the DDME-min1 parametrisation.  
}
\br
&\multicolumn{3}{l}{SLy5-min}  & & \multicolumn{3}{l}{DDME-min1}  \\
\cline{2-4}\cline{6-8}
  $A$    &   $A_0$&  & $\sigma(A_0)$ & & $A_0$ & &$\sigma(A_0)$ & units\\
\mr
& & & &  & & &  \\
SNM & &  & & & & &\\
\mr
  $\rho_0  $&$   0.162$ & $\pm$ & $0.002$  & &$   0.150$ & $\pm$ & $0.001$  &  fm${}^{-3}$\\    
  $e(\rho_0) $&$ -16.02$  & $\pm$ & $0.06$ & &$ -16.18$  & $\pm$ & $0.03$ &  MeV\\            
  $m^*/m $&$   0.698$ & $\pm$ & $0.070$    & &$   0.573$ & $\pm$ & $0.008$    &  \\               
  $J $&$  32.60$  & $\pm$ & $0.71$         & &$  33.0$  & $\pm$ & $1.7$         &  MeV\\    
  $K_0 $&$ 230.5$   & $\pm$ & $9.0$        & &$ 261$   & $\pm$ & $23$        &  MeV\\               
  $L $&$  47.5$   & $\pm$ & $4.5$         &  &$  55$   & $\pm$ & $16$         &  MeV\\                
& & & &  & & &  \\
${}^{208}$Pb & & & & & & & \\
\mr
  $E_x^{\rm ISGMR}$ &$  14.00$  & $\pm$ & $0.36$   & &$  13.87$  & $\pm$ & $0.49$   &  MeV\\               
  $E_x^{\rm ISGQR}$ &$  12.58$  & $\pm$ & $0.62$   & &$  12.01$  & $\pm$ & $1.76$   &  MeV\\                
  $\Delta r_{np} $&$   0.1655$& $\pm$ & $0.0069$ & &$   0.20$& $\pm$ & $0.03$ &  fm\\                  
  $E_x^{\rm IVGDR}$ &$  13.9$   & $\pm$ & $1.8$    &  &$  14.64$   & $\pm$ & $0.38$    &  MeV\\              
  $m_{-1}^{\rm IVGDR}$ &$  4.85$  & $\pm$ & $0.11$   & &$  5.18$  & $\pm$ & $0.28$   & MeV${}^{-1}$ fm$^2$\\
  $E_x^{\rm IVGQR}$ &$   21.6$  & $\pm$ & $2.6$     & & $25.19$ & $\pm$&$2.05$&   MeV    \\
\br
\end{tabular}
\end{indented}
\end{table}

By employing covariance analysis, statistical uncertainties related to SLy5-min and DDME-min1 parametrisations are calculated for several quantities of interest. Table~\ref{tabl2} shows the calculated nuclear matter properties, and a set of quantities for $^{208}$Pb, neutron skin thickness, centroid excitation energies of IVGDR, IVGQR, ISGMR, ISGQR and inverse energy weighted sum rule for IVGDR $(m_{-1})$. The respective calculated uncertainties are also shown. Theoretical errors appear relatively small (below 1$\%$) for a number of nuclear matter properties, including the saturation density $\rho_0$, corresponding binding energy $e(\rho_0)$ and effective masses. For the case of DDME-min1, the symmetry energy at saturation density $S_2(\rho_0)$, its slope $L$ and nuclear matter incompressibility $K_0$ result in relatively large uncertainties. This result is closely related to the fitting protocol employed, that is based only on the properties of finite nuclei. Large uncertainties in $S_2(\rho_0)$, $L$, and $K_0$ indicate that additional input related to nuclear matter properties seems to be necessary in the fitting procedure in order to provide improved constraints on the isovector channel of the energy density functional and related quantities such as symmetry energy parameters and neutron skin thickness. In the case of Skyrme functional, where additional nuclear matter constraints have been explicitly included in $\chi^2$ minimization (Sec.~\ref{skyrme}), the uncertainties in $S_2(\rho_0)$, $L$, and $K_0$ appear indeed smaller (Tab.~\ref{tabl2}). Considering the properties of excitations, statistical uncertainties in $E_x$ are in the range of $\approx1-3\%$ and, in the case of $m_{-1}$(IVGDR), $\approx 5\%$.

\subsection{Sensitivity of the $\chi^2$ definition on the predicted correlations}
\label{results2}

\begin{figure}[t!]
\includegraphics[clip=true,width=0.48\linewidth]{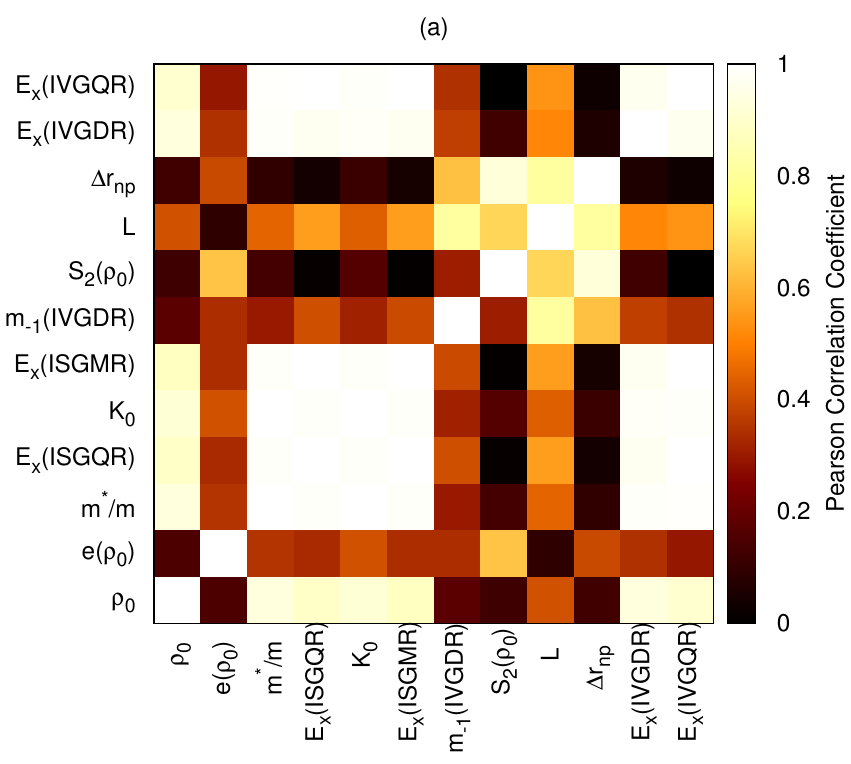}\hfill
\includegraphics[clip=true,width=0.48\linewidth]{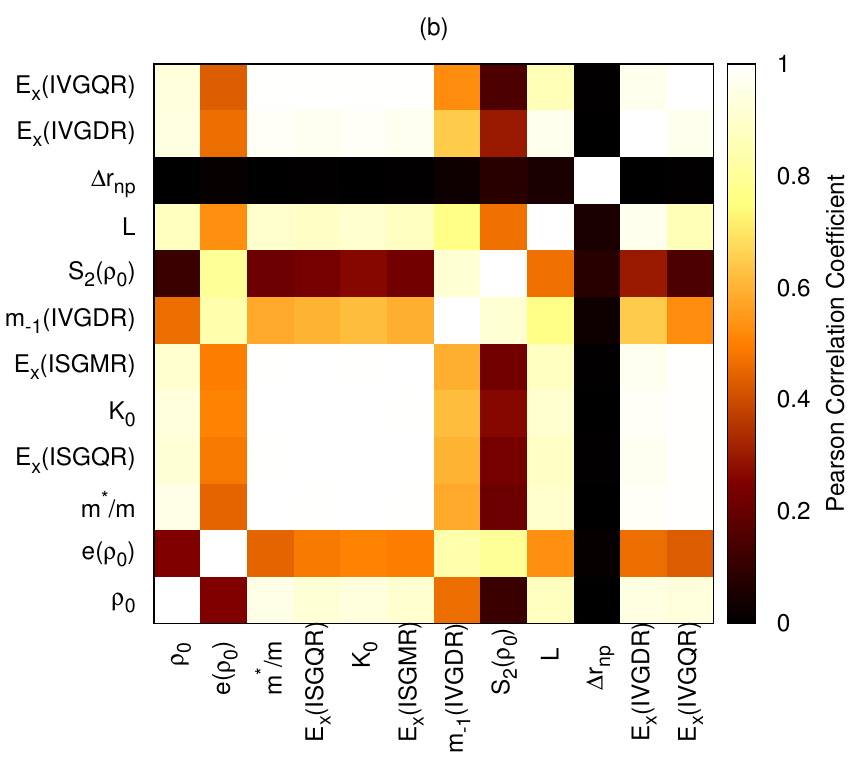}
\caption{Pearson product-moment correlation coefficient matrix (colour code) as predicted by the covariance analysis of two variants of the SLy5-min functional for different properties (see text for definitions and explanation on the two variants). Panel (a): SLy5-a. Panel (b): SLy5-b. \label{fig02}} 
\end{figure}

In this Section, we will try to clarify some of the {\it fake} puzzles briefly discussed in the previous section regarding the SLy5-min functional. Specifically, we will concentrate on those related to the neutron matter equation of state and the neutron skin thickness in ${}^{208}$Pb. For this purpose, we will use slightly different definitions of the $\chi^2$. Specifically, we have constructed two variants of SLy5-min. In SLy5-a we have kept all terms in the $\chi^2$ as in SLy5-min but we have changed that associated with the equation of state of neutron matter \cite{wiringa}. We have increased the value of $\Delta e(\rho,\delta=1)$ from $0.1\times e(\rho,\delta=1)$ --- that corresponds to a 10\% relative error --- to $0.5\times e(\rho,\delta=1)$. The Pearson-product correlation coefficients of this fit are shown in Fig.~\ref{fig02}.a where now the neutron radius of ${}^{208}$Pb appears to display a higher correlation with $S_2(\rho_0)$, $L$ and $m_{-1}$(IVGQR). This result clearly indicates that {\it when a constraint on a property is relaxed, correlations of other related observables not included in the fitting protocol with such a property should become larger.}

The second model we have built is named SLy5-b. In this case, we have kept all terms in the $\chi^2$ as in SLy5-min except the equation of state of neutron matter that now is not employed, and we used instead a very tight constraint on the neutron skin thickness of ${}^{208}$Pb: we have chosen as a test value $\Delta r_{np} = 0.160 \pm 0.001$ fm. Figure~\ref{fig02}.b confirms the expected result: $\Delta r_{np}$ display an almost zero correlation with all the other quantities\footnote[8]{Note that other isovector properties not tightly constrained in the fit appear mutually well correlated}. This is because there is not enough parameter space to explore variations on the $\Delta r_{np}$ in this example. This indicates that {\it when a property is tightly constrained --- artificially or by an accurate experimental measurement --- correlations of other observables with such a property should become small.}  

\section{Instabilities}
\label{instabilities}

Instabilities can impair the possibility to fit a new functional or to perform a sound correlation analysis or, generally speaking, to consider a functional as fully reliable. Instabilities in a functional can manifest themselves in several different ways. It is still unclear if there is a straightforward relationship between different kinds of instabilities. In general, we define as instability a situation in which a system described by a functional, when subject to some sort of perturbation, displays a divergent  or physically unreasonable behaviour. Recently there has been much interest in this topic, though mainly in connection with Skyrme functionals. Analysis of instabilities associated to covariant functionals are more sparse and less recent.

It is of course easier to detect instabilities in uniform matter. If we deal with a perturbation that transfers zero momentum (${\vec q}=0$), namely is characterised by infinite wavelength, we are in the so-called Landau limit. This case can be well described within the Landau's theory of Fermi liquids, that has been extended by Migdal and collaborators to the case of finite systems like nuclei \cite{Migdal:1967}. In the Landau-Migdal's theory the key quantity is the interaction potential $V$ acting among quasi-particles around the Fermi surface, whose matrix elements can be written in terms of the so-called Landau parameters $F,F',G$ and $G'$. 

In order for a spherical Fermi surface to be stable against any deformation, the parameters must satisfy the criterion
\begin{equation}
F_l>-(2l+1),
\label{eq:LM3}
\end{equation}
$l=0,1$ for s- and p-wave interactions, respectively. Analogous criteria holds for all the other parameters. For the standard Skyrme forces, only $l=0,1$ Landau parameters do not vanish and need to be considered. This is not the case for finite-range interactions.

Specific considerations are in order if a tensor force is added on top of the central terms, in the Landau-Migdal framework~\cite{Dab76, Back79,Olsson04}. In this case, due to the coupling between orbital angular momentum and spin, one must generalise the perturbing fields and impose that the Fermi surface is stable under the corresponding deformations that have total angular momentum and parity $J^\pi$ as quantum numbers. The resulting stability conditions that generalise Eq. (\ref{eq:LM3}) are written in Refs. \cite{Back79,Gang10}. A systematic study of the stability of a large set of Skyrme forces plus tensor terms has been carried out in Ref. \cite{Gang10}. One of the conclusions, that is of interest for the current paper, is that a full variational procedure to determine the Skyrme parameters is preferable to a perturbative adding of the tensor terms. 

Immediately afterwards, the question has been raised above the finite-$\vec q$ instabilities in uniform matter. To explore them, the Lyon group \cite{Davesne:2009,Davesne:2011} has developed a general response function formalism for a Skyrme force including central, spin-orbit and tensor terms. This work has generalised the previous works of Refs. \cite{Garcia-Recio:1992,Margueron:2006}. The response function of uniform matter is labelled by the indices corresponding to the total spin and isospin ($S$ and $T$), as well as by those corresponding to their projection on the quantisation axis ($M_S$ and $M_T$). The quantisation axis is chosen in the direction of the transferred momentum $\vec q$. The label $\alpha$ is chosen to denote the set ($S$, $M_S$; $T$, $M_T$). To find the response function $\chi^{(\alpha)}(q,\omega)$ one must solve the Bethe-Salpeter equation and obtain the RPA Green's function. From it,
the strength function $S^{(\alpha)}(q,\omega)$ is easily deduced.

The work of Ref. \cite{Davesne:2009} has been extended to a functional (not necessarily derived from a Hamiltonian) in \cite{Pastore:2012a} (cf. also Ref. \cite{Pastore:2014}). One of the main goals of this latter paper is making the detection of instabilities more efficient. In fact, if instabilities manifest themselves through an eigenvalue that crosses the zero value on the real axis, and evolves in the complex plane, the associated inverse-energy weighted sum rule $m_{-1}$ will have a pole. Seeking such poles is quite fast since the $m_{-1}$ sum rule possesses an analytical expression. 

\begin{figure}[htb]
\centering
\includegraphics[width=0.4\textwidth]{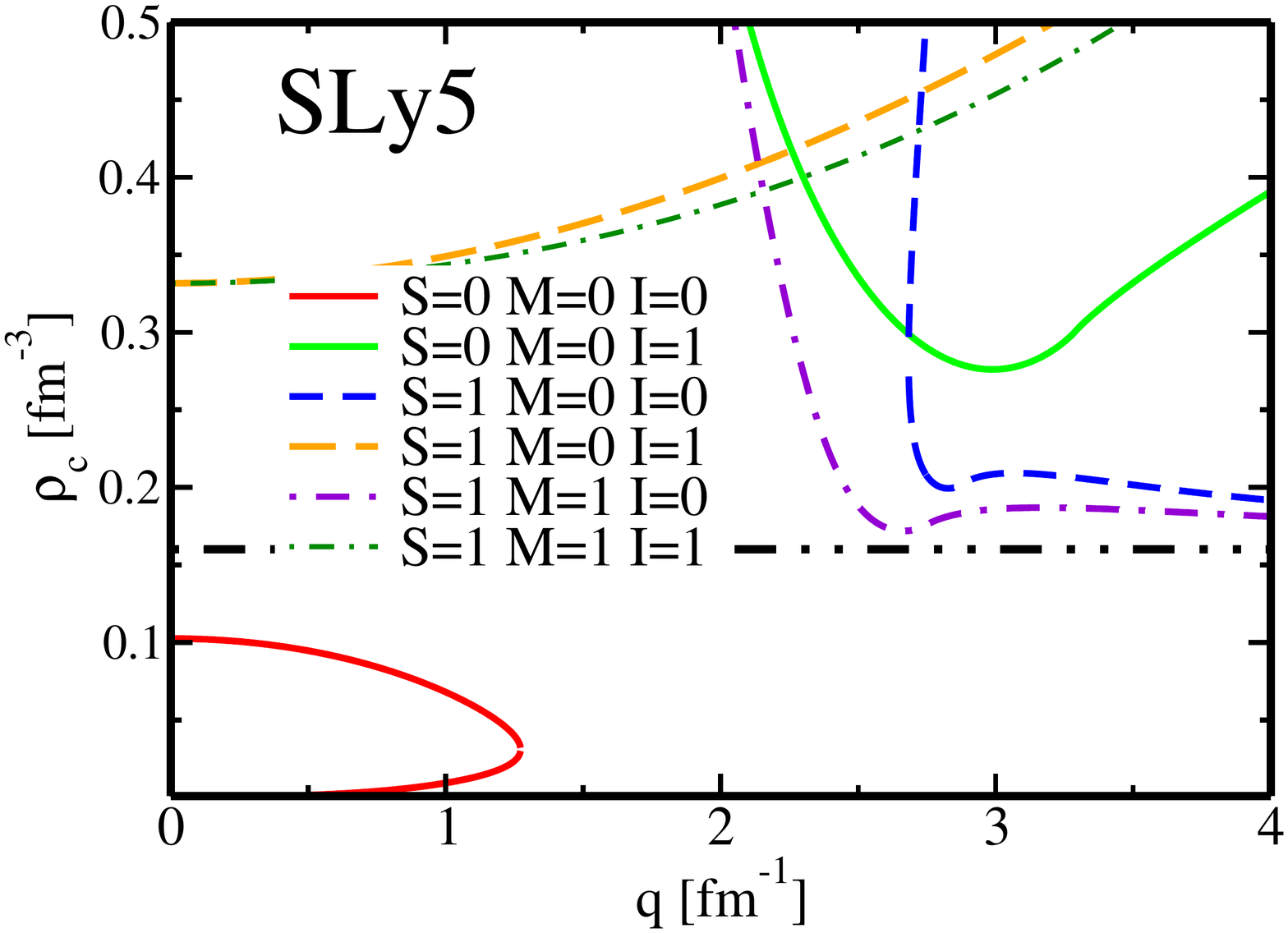}
\includegraphics[width=0.4\textwidth]{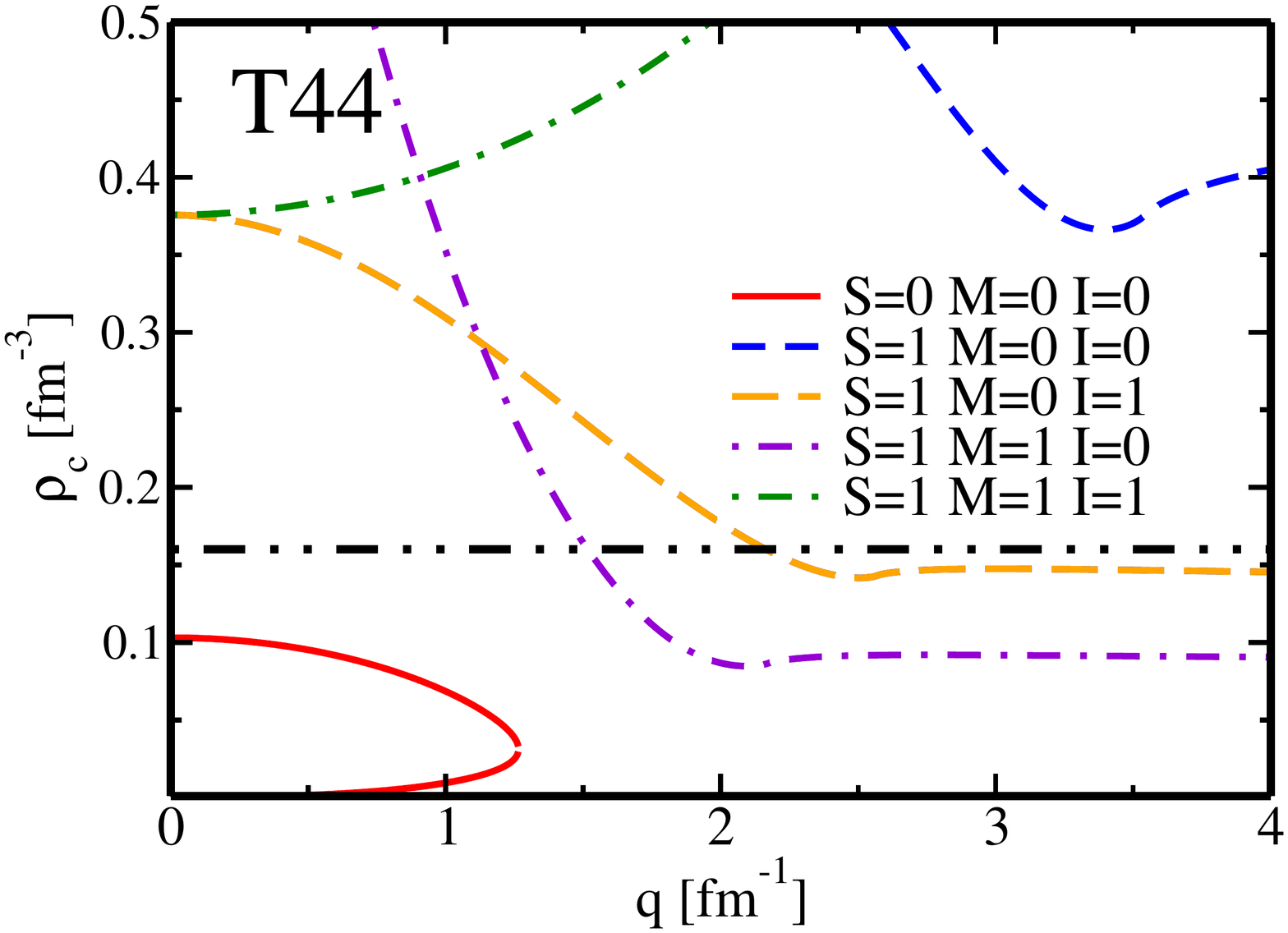}
\caption{\label{fig:instab1}
Critical densities $\rho_C$ as functions of the transferred momentum $q$, in symmetric nuclear matter and for the case of two Skyrme functionals that either do not include or include tensor terms \cite{Pastore}. They are displayed for different channels, and the saturation density is highlighted with a horizontal line. Figures taken from Ref.~\cite{Sagawa14}.}
\end{figure}

In this way, it has been found that a very large number of Skyrme functionals are plagued by instabilities. These can be either mechanical, spinodal instabilities (i.e. those in which the system is unstable against phase separation) associated with the $S=0, T=0$ channel, or spin and spin-isospin (i.e. ferromagnetic) instabilities. The presence of tensor terms favours, generally speaking, the rise of instabilities. This is illustrated in Fig. \ref{fig:instab1}. Instabilities take place in the different channels at a critical density $\rho_c$ for each value (zero or finite) of $q$. As a rule, the critical density may be lower in the case of finite $q$ than in the case $q=0$, that is, in the Landau limit described at the start of this Section. Intuitively, whereas the $q=0$ instability can be thought to concern the bulk medium as a whole, the finite-$q$ instability is a finite-size one taking place in a domain whose scale is $\Delta R \approx 2\pi/q$. In principle, this could be tolerable if the momentum scale (the real space scale $\Delta R$) is much larger (much smaller) than the typical low-energy nuclear physics scale. Thus, the question about a maximum $q$ and a maximum $\rho_c$ at which instabilities are acceptable, should be asked. Some groups are at present developing fitting protocols of a Skyrme functional in which the requirement that no instability should be present, at least at densities $\lesssim$ 1.2 times the saturation density, is enforced (except for spinodal instabilities that are believed to have a physical meaning and that, anyway, take place at lower densities than those of interest for nuclear structure).

In Ref. \cite{Navarro} an interesting comparison between instabilities displayed by either zero-range or finite-range interactions has been carried out. It has been confirmed that in the case of zero-range interactions the addition of tensor terms favours the appearance of instabilities, but this is not the case for the finite-range forces. For instance, the force M3Y-P2 \cite{Nakada1} is quite free from instabilities although containing a genuine tensor part. We should also keep in mind that ferromagnetic instabilities displayed by effective interactions do not appear in {\it ab-initio} calculations of uniform matter. The authors of \cite{Navarro} have calculated the spin susceptibility $\chi_{\rm RPA}(0)$. The trend of this quantity, as predicted by realistic forces, is better followed by finite-range forces than by zero-range forces.

We move to a discussion of the instabilities that appear in calculations of finite nuclei. One of the earliest findings, in this respect, has been that in Skyrme Hartree-Fock calculations for standard double magic nuclei, after a sufficiently long number of iterations the system converges to an unphysical state in which proton densities and neutron densities are separated apart \cite{Lesinski:2006}. Another case of instability occurred in cranked-HFB calculations performed in $^{194}$Hf, where it has been found that the system was sometimes converging to a spin-polarised state \cite{Hellemans:2012}. The most recent analysis can be found in Ref. \cite{Hellemans:2013}: there are still uncertainties in relating instabilities in finite and infinite systems and one should keep in mind, on top of this, that the numerical scheme used for finite systems does actually play a significant role. 

We end this Section by considering the case of instabilities in relativistic functionals. It must be stressed that the knowledge on instabilities in the relativistic framework is less systematic and rather limited, being based mainly on the Walecka $\sigma-\omega$ model with finite-range meson exchange, and its extensions with nonlinear scalar and vector self-interaction and scalar-vector coupling terms. Therefore, most of the early investigations on instabilities should be repeated with more advanced relativistic EDFs, in particular those with density dependent meson-nucleon couplings. In principle, the covariant framework allows exploring more instabilities than the nonrelativistic one; nonetheless, there are some instabilities that parallel those seen above. 

Some attention has been paid to instabilities due to quantum fluctuations. Studies of uniform nuclear matter have been made more or less at the same time in \cite{Furnstahl:1988} and \cite{Friman:1988}, and slightly later in \cite{Dobereiner:1989} (see also \cite{Henning:1988}). In the first two works the Walecka  $\sigma-\omega$ model is employed, while Ref. \cite{Dobereiner:1989} deals with the model including non-linear self-couplings of the scalar $\sigma$ field. As pointed out clearly in \cite{Furnstahl:1988}, the vacuum polarisation instabilities take out at rather large momenta; however, in nuclear matter the instabilities coming from p-h insertions in the meson propagator can appear at lower $q$. Interestingly, this is another case in which one can compare with the spin instabilities that have been mentioned above in the context of Skyrme functionals: in fact, in Ref. \cite{Friman:1988} it has been shown by means of a nonrelativistic reduction of the transverse part of $\omega$-exchange that the state resulting from the instability is a spin-polarised state. 

Spinodal instabilities take place at higher densities in the Walecka model than in the Skyrme case, as it has been shown in \cite{Ayik:2009}. However, this seems to be a specific feature of that model, as both $\sigma-\omega$ models with non-linear terms and the DD-ME1 Lagrangian display a behaviour that resembles that of the Skyrme forces \cite{Ayik:2011}. In these works, the relativistic transport (i.e. Vlasov) equations have been solved in a semiclassical framework. There are also works addressing spinodal and other instabilities using the response function formalism in the relativistic mean field models, extended with nonlinear self-interaction terms of the $\sigma$-meson and $\omega$-meson fields, as well as with nonlinear vector-scalar terms \cite{Sulaksono:2006,Sulaksono:2007}. Finally, specific instabilities taking place in the environment inside neutron stars, when such kind of matter is studied by means of models that include the $\delta$-meson field, are addressed in \cite{Rabhi:2009,Pais:2009}. We are not aware of any paper devoted to instabilities when the relativistic framework is employed in description of finite nuclei, at variance with the Skyrme case.

\section{Conclusions}
\label{conclusions}

Most of available nuclear energy density functionals omit the theoretical estimation of errors and correlations between parameters and computed quantities. In this contribution we highlight the relevance of performing the covariance analysis in order to assess the information content of an observable. Such an analysis provides an estimation of the {\it statistical uncertainties} and correlations associated to any predicted quantity on the basis of the experimental data used for defining the quality measure. It is important to note that other sources of theoretical errors exist though they are not the focus of the present contribution \cite{erler12,bevington,dobaczewski14}. 

We have briefly presented the formalism of covariance analysis and discussed the results of two successful nuclear energy density functionals: a non-relativistic Skyrme functional built from a zero-range effective interaction; and a relativistic nuclear energy density functional based on density dependent meson-nucleon couplings. The covariance analysis of these models has allowed us to provide meaningful statistical errors in the parameters (Table~\ref{tabl1}) and in some predicted observables (Table~\ref{tabl2}). As it may be expected, we have seen that the errors calculated for the different nuclear properties appear to be relatively small for a number of properties that are known to be well constrained by the employed experimental data defining the quality measure while large errors are found when the defined quality measure lacks the data needed to constrain them. {\it A large error for a given non-fitted observable indicates that the quality measure does not contain enough related information}. The solution to this problem is to inspect the set of data used for adjusting the parameters and try to optimise it.    

We have also studied in some detail the correlations displayed by a set of selected nuclear properties. An overall picture in which most of the strongest correlations are between properties of either isoscalar or isovector nature separately is given. However, a more careful analysis of the results indicates that the picture is not always so clear. We have shown that in some of the cases, one needs some physical understanding in order to unveil the origin of some of the correlations --- or lack of correlation --- between some of the analysed observables. Therefore, {\it some useful insights on the physical understanding of the system under study might be fostered by a simple covariance analysis after the optimal parametrisation of the model is determined}. 

In order to investigate and show the relevance of the definition of the quality measure and its impact on the optimised model, we have explored two variants of the SLy5-min functional in which the $\chi^2$ has been slightly modified. In the first example, the weight of the neutron matter equation of state is relaxed. These results clearly indicate that {\it when a constraint on a property is released, correlations of other related observables not included in the fitting protocol with such a property should become larger.} In the second example, the weight of the neutron matter equation of state is further relaxed and the $\Delta r_{np}$ in ${}^{208}$Pb is added into the $\chi^2$ definition with a very small adopted error. The results are transparent, namely there is not enough parameter space to explore variations on the $\Delta r_{np}$. {\it When a property is tightly constrained --- artificially or by an accurate experimental measurement --- correlations of other observables with such a property should become small.}   

Finally, instabilities should be avoided if one wants to build a reliable energy density functional. The only physically known instability is the spinodal instability at low densities, while other instabilities like in particular the spin or spin-isospin ones do not show up in any calculation using realistic interactions. Of course, the discussion about the regime in which one can tolerate instabilities, namely the maximum momentum values, is strictly related to the more general discussion about the momentum scale in which these functionals can be applied, in a sort of effective field theory spirit. While non-relativistic models have been more carefully studied, and groups are starting to insert procedures to avoid instabilities in their protocols, the situation is far less clear in the relativistic case. Certainly one could guess that finite-range models are less prone to instabilities than zero-range. However, as density-dependent point-coupling models are becoming increasingly popular, the issue deserves further investigation. Of particular importance are studies based on relativistic point coupling models, which due to the zero-range interaction would provide the insight into relationships between instabilities in non-relativistic (Skyrme) and relativistic models.

\ack
We are indebted to K. Bennaceur, A. Pastore, P.-G. Reinhard, and P. Ring for valuable discussions and useful correspondence. N.P. acknowledges support from the Cooperation programme of the University of Zagreb.

\section*{References}

\bibliography{bibliography}

\end{document}